\documentclass[aps,prl,twocolumn,amsmath,amssymb,tightenlines,epsfig,floatfix]{revtex4}
\usepackage{graphicx}
\usepackage{dcolumn}	
\usepackage{bm}			
\usepackage{amsfonts}
\usepackage{xspace}

\begin{document}

\newcommand{\psihat}{\ensuremath{\hat{\psi}}\xspace}
\newcommand{\psihatd}{\ensuremath{\hat{\psi}^{\dagger}}\xspace}
\newcommand{\ahat}{\ensuremath{\hat{a}}\xspace}
\newcommand{\Ham}{\ensuremath{\mathcal{H}}\xspace}
\newcommand{\ahatd}{\ensuremath{\hat{a}^{\dagger}}\xspace}
\newcommand{\bhat}{\ensuremath{\hat{b}}\xspace}
\newcommand{\bhatd}{\ensuremath{\hat{b}^{\dagger}}\xspace}
\newcommand{\boldr}{\ensuremath{\mathbf{r}}\xspace}
\newcommand{\dr}{\ensuremath{\,d^3\mathbf{r}}\xspace}
\newcommand{\etal}{\emph{et al.\/}\xspace}
\newcommand{\ie}{i.e.\:}
\newcommand{\eq}[1]{Eq.\,(\ref{#1})\xspace}
\newcommand{\fig}[1]{Figure\,(\ref{#1})\xspace}
\newcommand{\abs}[1]{\left| #1 \right|} 

\title{Information recycling beam-splitters for quantum enhanced atom-interferometry. }
\author{S. A. Haine}
\affiliation{School of Mathematics and Physics,  University of Queensland, Brisbane, QLD, 4072, Australia.}
\email{haine@physics.uq.edu.au}

\begin{abstract}
We propose a scheme to significantly enhance the sensitivity of atom-interferometry performed with Bose-Einstein condensates. When an optical two-photon Raman transition is used to split the condensate into two modes, some information about the number of atoms in one of the modes is contained in one of the optical modes. We introduce a simple model to describe this process, and find that by processing this information in an appropriate way, the phase sensitivity of atom interferometry can be enhanced by more than a factor of 10 for realistic parameters. 
\end{abstract}

\maketitle

 Atom interferometers are devices that exploit the wave-like properties of atomic systems, and can provide sensitive inertial measurements \cite{chu99, chu2001, kasevich02, kasevich97}, as well as measurements of magnetic fields \cite{stamperkurn07}, the gravitational constant ($G$) \cite{kasevichG}, and the fine structure constant ($\alpha$) \cite{biraben_alpha}. Although most state-of-the-art atom interferometers currently utilise laser cooled thermal atoms, there are some benefits to using Bose-condensed atoms, as they provide improved visibility in configurations which require complex manipulation of the motional state such as high momentum transfer beamsplitters \cite{close2011}. The ultimate limit to the sensitivity of any interferometric device is the Heisenberg limit $\Delta \phi = \frac{1}{N_t}$, where $N_t$ is the total number of detected particles. Due to the linear nature of atomic beam splitters, almost all atom interferometers demonstrated so far operate with uncorrelated atoms in each arm. This puts a limit on the sensitivity of $\Delta \phi = \frac{1}{\sqrt{N_t}}$, which we refer to as the \emph{standard quantum limit} (SQL) \cite{dowling}. 

As it is technically challenging to increase the atomic flux in these devices, there is much interest in surpassing the SQL in atom interferometers via the use of quantum entanglement. Recently, there have been two proof of principle experiments demonstrating sensitivity beyond the SQL by generating spin-squeezing via one-axis twisting \cite{ueda, hainekerr} resulting from the nonlinearity induced by atomic collisions \cite{oberthaler2010, treutlein2010}. However, both of these experiments use only a small number of atoms ($200$-$1200$ atoms), so the absolute sensitivity of the device is low. Recently, correlated pairs of atoms generated from atomic spin-exchange collisions have been used to perform interferometry below the SQL \cite{ertmer2011}. The number of pair-correlated atoms used in the interferometer was approximately $8000$, about $1/4$ of the total available atoms in the experiment. 

In this letter, we propose a scheme for surpassing the SQL via a different approach. Instead of using a nonlinear atomic process to create entanglement between two atomic modes which are subsequently used as the input to an interferometer, we use the beam splitting process itself to create correlations between one of the atomic modes, and an optical field. The two atomic modes remain uncorrelated during the interferometer process, but by making appropriate measurements on the optical beam, a correction to the atomic signal is obtained that surpasses the SQL. The benefit of this scheme is that it does not rely on atomic interactions to create the correlations, so  can operate in a dilute regime where the detrimental effects of phase-diffusion and multimode excitations due to atomic interactions will be negligible. 

\begin{figure}
\includegraphics[width=0.4\columnwidth]{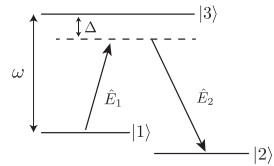}
\caption{\label{fig:scheme1} Energy level scheme for a three-level Raman transition comprising two nondegenerate hyperfine ground states ($|1\rangle$ and $|2\rangle$). The BEC is initially formed in state $|1\rangle$, and populations is transferred to $|2\rangle$ via the absorption of a photon from $\hat{E}_1$ and the emission of a photon into $\hat{E}_2$, both detuned from the excited state ($|3\rangle$) by an amount $\Delta$. }
\end{figure}

\emph{Scheme} --- Our atom interferometry scheme is based on interference between two, non-degenerate, hyperfine ground states ($|1\rangle$ and $|2\rangle$) of a Bose-Einstein condensate (BEC) comprised of three-level atoms (Fig. (\ref{fig:scheme1})). Atom-light correlations are created via an atomic beam splitter based on a Raman transition, in which states $|1\rangle$ and $|2\rangle$ are coupled by two optical fields $\hat{E}_1$ and $\hat{E}_2$ detuned from an excited state $|3\rangle$. When an atom is transferred from $|1\rangle$ to $|2\rangle$, a photon is absorbed from $\hat{E}_1$ and emitted into $\hat{E}_2$. In principle, by measuring the number of photons in $\hat{E}_2$, it may be possible to gain some information about the number of atoms transferred, and therefore the number difference between the two modes. However, when using a conventional Raman transition with bright beams, the atom-light correlations between modes $|2\rangle$ and $\hat{E}_2$ will be swamped by the large number of uncorrelated photons initially in $\hat{E}_2$. We can circumvent this by using a Raman super-radiance transition \cite{Schneble04, Yoshikawa04, Cola04, Wang05, Meystre07, Meystre2000,Hilliard08}, in which the initial population of $\hat{E}_2$ is set to zero, and the population transfer between $|1\rangle$ and $|2\rangle$ is driven by the absorption of a photon from $\hat{E}_1$, and the spontaneous emission of a photon into $\hat{E}_2$. A small `seed' of atoms in $|2\rangle$, which could be populated via a conventional Raman transition, stimulates the transfer such that the majority of photons are emitted into a single mode. The sequence of our scheme is as follows: We begin with a BEC of $N_t$ atoms all in mode $|1\rangle$. Immediately after a seed of $N_{seed}$ atoms is transferred via a Raman transition, the Raman-super-radiance step is implemented by switching on $\hat{E}_1$ from $t_0$ to $t_1$, which is chosen such that the desired amount of population transfer is reached. A signal $S_b$ is then obtained by measuring $\hat{E}_2$  via Homodyne detection with a local oscillator which is phase locked with $\hat{E}_1$ and the light used in the initial Raman transition to create the initial seed in $|2\rangle$. The two modes of atoms then undergo a conventional atom interferometry scheme: At $t_1$, a conventional $\frac{\pi}{2}$ Raman transition exchanges population between the two modes until $t_2$. A phase shift is then added to mode $|2\rangle$ caused by the physical process to be measured, before the two modes are interfered again via another conventional $\frac{\pi}{2}$ Raman transition. At $t_3$, a signal $S_a$ is obtained by measuring the difference in the number of particles in each mode, and using this to estimate the magnitude of the applied phase shift. By combining $S_a$ with $S_b$ in the appropriate way, the sensitivity can surpass the standard quantum limit. 

\begin{figure}
\includegraphics[width=0.8\columnwidth]{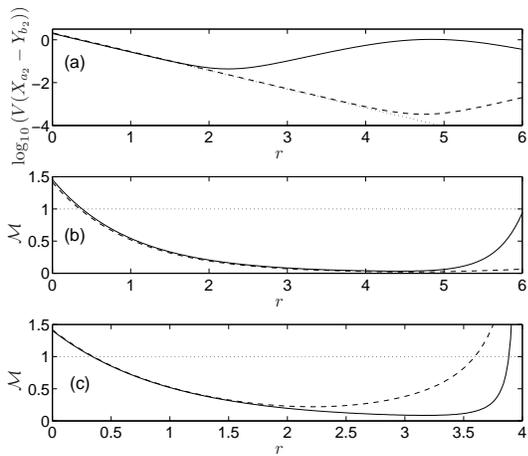}
\caption{\label{fig:var_vs_r}  (a): $\log_{10}\left(V(\hat{X}_{a_2}-\hat{Y}_{b_2})\right)$ at $t_1$ vs. $r$ calculated via the undepleted pump approximation (dotted trace), and from the TW model for $N_{seed}=0$ (dashed trace), and $N_{seed} =10^4$ atoms (solid trace). The total number of atoms was $N_t =10^7$. In the case with an initial seed, the point of maximum correlations corresponds to approximately $2.3\times 10^5$ atoms transferred during the Raman-super-radiance step. (b): $\mathcal{M}$ vs. $r$ for $N_{seed} =0$ from the undepleted pump approximation (dashed line), and the TW model (solid line). (c): $\mathcal{M}$ vs. $r$ for $N_{seed}=10^4$ from the undepleted pump approximation (dashed line), and the TW model (solid line). A value of $\mathcal{M}<1$ indicates sensitivity greater than the SQL. }
\end{figure}

\emph{Model} --- Expanding $\hat{E}_2(\boldr)$ in the plane-wave basis, $\hat{E}_2(\boldr) =  \sum_\mathbf{k} e^{i\mathbf{k}\cdot \boldr}\bhat_\mathbf{k}$, the Hamiltonian for the system is then
\begin{eqnarray}
\Ham &=& \sum_{i=1}^3 \int \psihatd_i(\boldr) H_i \psihat_i(\boldr) \, \dr +  \sum_\mathbf{k}\hbar c k \bhatd_\mathbf{k}\bhat_\mathbf{k} \nonumber\\
&+& \hbar\int \left( \Omega e^{i\left(\mathbf{k}_0\cdot\boldr - (\omega_0-\Delta)t\right)}\psihatd_3(\boldr)\psihat_1(\boldr) + h.c.\right)\dr \nonumber \\
 &+& \hbar\int\left(\psihatd_3(\boldr)\psihat_2(\boldr)\sum_{\mathbf{k}} g_\mathbf{k} \bhat_\mathbf{k}  e^{i \mathbf{k}\cdot \boldr}+h.c.\right)\dr \, ,
\end{eqnarray}
where $\psihat_i(\boldr)$ is the annihilation operator for atoms in electronic state $|i\rangle$, and we have assumed that $\hat{E}_1$ is sufficiently bright that we can ignore depletion and treat it semiclassically with Rabi frequency $\Omega$ and wavevector $\mathbf{k}_0$, with $c|\mathbf{k}_0| = \omega_0 - \Delta$.  We can capture the important physics of the system, in particular the quantum correlations between the atoms and $\hat{E}_2$ with a simplified toy model. Expanding the field operators as $\psihat_1(\boldr) = \sum_j u_{j}(\boldr)\ahat_{1,j}$ and $\psihat_2(\boldr) = \sum_j u_{j}(\boldr)e^{i(\mathbf{k}_0-\mathbf{k}_2)\cdot\boldr}\ahat_{2,j}$, and assuming that initially modes $\ahat_{1,0} \equiv \ahat_1$ and $\ahat_{2,0}\equiv \ahat_2$, are highly occupied, with all other modes unoccupied, it is a reasonable assumption that the stimulated scattering from $\ahat_{1}$ into $\ahat_{2}$ is much higher than the scattering into all other modes $\ahat_{2,j}$. We will investigate the validity of this approximation below. Ignoring all modes other than the highly occupied modes, and assuming that $\Delta$ is large enough to adiabatically eliminate the excited state, as in \cite{squeezy1, squeezy3}, the Heisenberg equations of motion become
\begin{eqnarray}
i\dot{\hat{a}}_1 &=&  - \hat{a}_2\sum_{\mathbf{k}} G_\mathbf{k} \tilde{b}_\mathbf{k} \label{a1dot} \, , \\
i\dot{\hat{a}}_2 &=&  - \hat{a}_1\sum_{\mathbf{k}} G_\mathbf{k}^* \tilde{b}^\dagger_\mathbf{k} \label{a2dot} \, ,\\
i \dot{\tilde{b}}_\mathbf{k} &=& \left(\omega_2 - c\left(|\mathbf{k}_0| - |\mathbf{k}|\right)\right)\tilde{b}_\mathbf{k} - G_{\mathbf{k}}^* \ahat_1\ahatd_2 \label{b2dot} \, ,
\end{eqnarray}
where $\tilde{b}_\mathbf{k} \equiv \bhat_\mathbf{k} e^{i(\omega_0-\Delta -\omega_2)t}$, $\tilde{a}_2 = \ahat_2e^{i\omega_2 t}$, $G_\mathbf{k} \equiv \frac{g_\mathbf{k}^* \Omega}{\Delta}\int |u_0(\boldr)|^2 e^{i(\mathbf{k}-\mathbf{k}_2)\cdot\boldr}\dr$,  $H_i u_0(\boldr) = \hbar\omega_i u_0(\boldr)$, and we have set our zero of energy such that $\omega_1 =0$. In writing \eq{a2dot} we have assumed that the timescale for state $|2\rangle$ atoms to move due to their momentum kick is large compared to the dynamics of the population transfer. If the characteristic size of the condensate is much larger than the optical wavelength, then $G_\mathbf{k}$ will be sharply peaked around $\mathbf{k}=\mathbf{k}_2$ (which corresponds to $ \omega_2 - c\left(|\mathbf{k}_0| - |\mathbf{k}|\right) =0$), so it is reasonable to treat the optical field as a single mode $\bhat_2$, which corresponds to the optical mode which conserves momentum and energy when transferring atoms from $\ahat_1$ to $\ahat_2$. We can gain an understanding of how the atom-light entanglement enhances the interferometry by assuming that the number of atoms transferred to $|2\rangle$ during the Raman super-radiance step is a small fraction of the total number of atoms, allowing us making the undepleted pump approximation for $|1\rangle$, ie $\hat{a}_1 \rightarrow \sqrt{N}_1$, in which case the solution to \eq{a2dot} and \eq{b2dot} is  $\tilde{a}_2(t_1) = \tilde{a}_2(t_0)\cosh r -i \tilde{b}_2(t_0)\sinh r$ and $\tilde{b}_2(t_1) = \tilde{b}_2(t_0)\cosh r -i \tilde{a}_2(t_0)\sinh r$ \cite{Walls}, where  $r \equiv \sqrt{N}_1G_{\mathbf{k}_2} (t_1-t_0)$. We then implement the atom interferometer via a pair of conventional $\frac{\pi}{2}$ Raman transitions, such that $\tilde{a}_2(t_2) = \frac{1}{\sqrt{2}}\left( \tilde{a}_2(t_1) -i \ahat_1(t_1)\right)$, and $\ahat_1(t_2) = \frac{1}{\sqrt{2}}\left( \ahat_1(t_1) -i \tilde{a}_2(t_1) \right)$, $\tilde{a}_2(t_3) = \frac{1}{\sqrt{2}}\left( \tilde{a}_2(t_2)e^{i\phi} -i \ahat_1(t_2)\right)$, and $\ahat_1(t_3) = \frac{1}{\sqrt{2}}\left( \ahat_1(t_2) -i \tilde{a}_2(t_2)e^{i\phi} \right)$. The sensitivity is given by $\Delta \phi = \sqrt{\frac{V(S)}{(d\langle S\rangle /d\phi)^2}}$, where $S$ is the quantity measured, usually the number difference in the two modes, and $V(S) = \langle S^2\rangle - \langle S\rangle^2$.  If we chose the signal to be $S =S_a \equiv(\tilde{a}_2^\dag(t_2)\tilde{a}_2(t_2) - \ahatd_1(t_2)\ahat_1(t_2))$, then the point where $|d \langle S\rangle /d\phi|$ is maximum (and hence the most sensitive point) is $\phi =\frac{\pi}{2}$. In this case, when the number of atoms transferred in the super-radiance step is small compared to the total number of atoms, \ie, $\sinh^2 r \ll N_t$, we can think of $S_a$ as approximating a homodyne measurement of  the amplitude quadrature of $\tilde{a}_2(t_1)$, ie $S_a \approx-\sqrt{N}_t\hat{X}_{a_2}$, Where $\hat{X}_{a_2} =   \tilde{a}_2(t_1)+ \tilde{a}_2^\dag(t_1)$. Assuming the initial state of $|2\rangle$ is vacuum, then $\Delta \phi \approx \sqrt{\frac{\cosh 2r}{N_t}}$, which is greater than the SQL for all $r>0$. However, we can enhance our signal by measuring the scattered photons by mixing $\tilde{b}_2(t_1)$ with a strong local oscillator ($\bhat_{LO}$) assumed to be a coherent state $|\beta_{LO}\rangle$ on a 50/50 beam splitter, such that $\hat{c} = \frac{1}{\sqrt{2}}\left( \tilde{b}_2 -i\bhat\right)$, $\hat{d} = \frac{1}{\sqrt{2}}\left(\bhat - i\tilde{b}_2\right)$. By taking our signal to be $S = S_a - \frac{1}{g}S_b$, where $S_b = \hat{c}^\dag\hat{c} - \hat{d}^\dag\hat{d} \approx - \beta_{LO}\hat{Y}_{b_2}$ where $\hat{Y}_{b_2} = i\left(\tilde{b}_2(t_1) - \tilde{b}_2^\dag(t_1)\right) $ is the phase quadrature of $\hat{b}_2$, and $g = \frac{\beta_{LO}}{\langle \ahat_1(t_1)\rangle}$ is the gain factor, which is the square root of the ratio of the occupations of the local oscillator and condensate. In this case the maximum sensitivity (again, at $\phi = \frac{\pi}{2}$) is $\Delta \phi \approx \frac{\sqrt{2}e^{-r}}{\sqrt{N_t}}$, which will surpass the SQL for $r > \ln \sqrt{2} \equiv r_{crit}$. The reason for this enhancement is simple: As the atom-interferometer can be thought of as approximating an amplitude quadrature measurement of $\ahat_2(t_1)$, and our simple undepleted pump approximation for the Raman super-radiance leads to reduced fluctuations in $\left(\hat{X}_{a_2}-\hat{Y}_{b_2}\right)$ for $r> r_{crit}$ (Figure (\ref{fig:var_vs_r}) (a)), subtracting these two quantities will remove some of the quantum fluctuation which contribute to the noise. For $r<r_{crit}$, subtracting $\frac{1}{g}S_b$ from the signal detracts from the sensitivity, as we are essentially adding an extra source of uncorrelated noise. Figure (\ref{fig:var_vs_r}) (b) and (c) shows $\mathcal{M} \equiv \Delta \phi\sqrt{N_t}$ vs. $r$ for the simple case of $N_{seed}=0$, and $N_{seed} =10^4$ added to suppress the spontaneous emission into the other modes. In both (b) and (c), the minimum value of $\mathcal{M}$ is determined from the point when the atomic signal $S_a$ begins to deviate significantly from a perfect homodyne measurement of the amplitude quadrature of $\tilde{a}_2(t_1)$, due to the finite number of particles in $\hat{a}_1$. 

\begin{figure}
\includegraphics[width=0.95\columnwidth]{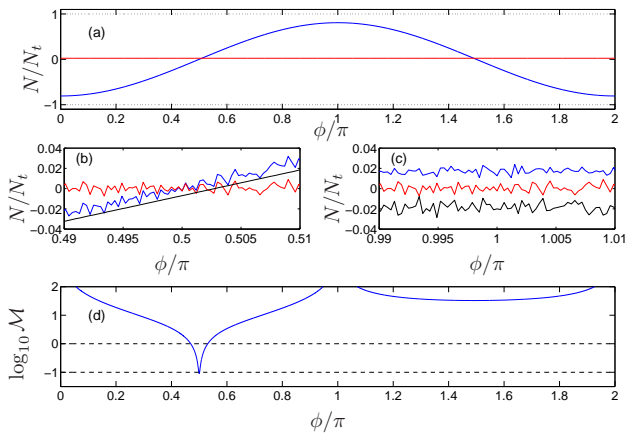}
\caption{\label{fig:merrit_vs_phi} (a) $\langle S_a\rangle$ (blue) and $\langle S_b\rangle$ (red) vs. $\phi$. (c) $S_a$ (blue), $\frac{1}{g} S_b$, (red), and $S\equiv S_a-\frac{1}{g} S_b$ (black) vs. $\phi$. Each point is a seperate trajectory from the stochastic simulation. A vertical offset has been added to some of the traces for the purpose of visual clarity. (d) $\log_{10}\mathcal{M}$ vs $\phi$. Values of $\log_{10}\mathcal{M} <0$ indicate surpassing the SQL. The dashed lines are to guide the eye.}
\end{figure}

\emph{Effects of depletion from the condensate} --- The undepleted pump approximation breaks down when the number of atoms transferred during the super-radiance step becomes significant, which is the case when an initial seed is used. We can model the effects of depletion from $\ahat_1$ via an approximate stochastic phase space method. Specifically, we use the truncated Wigner (TW) approach \cite{Walls, Steel98}, where Eqs. \ref{a1dot}, \ref{a2dot}, and \ref{b2dot} are converted into a set of stochastic differential equations, which we solve numerically.  By averaging over many trajectories (in this case, $1000$) with initial states sampled from the Wigner distribution, expectation values of quantities involving symmetrically ordered quantum field operators can be calculated \cite{Olsen09}. The stochastic differential equations are
\begin{eqnarray}
i\dot{\alpha}_1 &=& -G_{\mathbf{k}_0} \beta_2 \alpha_2  \label{alpha1dot} \\
i\dot{\alpha}_2 &=& -G^*_{\mathbf{k}_0} \alpha_1\beta_2^* \label{alpha2dot} \\
i\dot{\beta}_2 &=& -G_{\mathbf{k}_0}^* \alpha_1 \alpha_2^* \label{beta2dot} 
\end{eqnarray}
where we have made the operator correspondence $\ahat_1 \rightarrow \alpha_1$, $\tilde{a}_2 \rightarrow \alpha_2$, and $\tilde{b}_2 \rightarrow \beta_2$. We solved Equations \ref{alpha1dot}, \ref{alpha2dot}, and \ref{beta2dot} numerically with initial conditions consistent with Glauber coherent states \cite{Walls} with mean occupations of $\langle \ahatd_1\ahat_1\rangle = N_t - N_{seed}$, $\langle \ahatd_2\ahat_2\rangle = N_{seed}$, and $\langle \bhatd_2\bhat_2\rangle = 0$.  Figure (\ref{fig:var_vs_r}) (b) and (c) shows $\mathcal{M}$ vs. $r$. When $N_{seed}=0$, the maximum sensitivity is $\mathcal{M} \approx 0.03$, or equivalent to approximately $1000$ times more atoms. When $N_{seed}=10^4$, the maximum sensitivity is $\mathcal{M} \approx 0.09$, or equivalent to approximately $120$ times more atoms. These maximums occur when the number of atoms transferred during the super-radiance step is $2000$ and $10^6$ respectively. It is interesting to note that in this case, the TW model predicts \emph{more} quantum enhancement than the undepleted pump approximation, even though there is significantly less entanglement between $\tilde{a}_2(t_1)$ and $\tilde{b}_2(t_1)$. The reason for this is that the depletion from the condensate causes anti-correlations in the population between modes $\hat{a}_1$ and $\tilde{a}_2$ which contribute favourably to the signal. As a check, we replaced $\hat{a}_1$ with a coherent state uncorrelated with $\tilde{a}_2$ and recovered a sensitivity slightly less than that predicted by the undepleted pump approximation, consistent with level of the entanglement present. Fig. (\ref{fig:merrit_vs_phi}) (a) shows $\langle S_a\rangle$, and $\langle S_b\rangle$ vs. $\phi$ for $r=3$. Fig. (\ref{fig:merrit_vs_phi}) (b) and (c) show the value of $S_a$, $\frac{1}{g}S_b$  and $S$ as a function of $\phi$ for an individual realisation (each point corresponds to an individual trajectory of our stochastic simulation). In (b), for phases close to $\phi = \frac{\pi}{2}$, we see that the fluctuations in $\frac{1}{g}S_b$ are highly correlated with the fluctuations in $S_a$, and  the fluctuations in $S$ are almost completely removed. In (c), for phases close to $\phi = \pi$, the fluctuations are completely uncorrelated, and $S$ is strongly fluctuating. For phases close to $\phi = \frac{3}{2}\pi$, $S_a$ and $\frac{1}{g}S_b$ are strongly \emph{anti}-correlated, and we could recover a `quiet' $S$ by simply switching the sign of $g$. \fig{fig:merrit_vs_phi} (d) shows $\mathcal{M}$ vs. $\phi$. At $\phi = \frac{\pi}{2}$, $\mathcal{M} \approx 0.09$, significantly lower than the limit set by the SQL: $\mathcal{M}=1$. For this simulation, we chose $g=100$.

It should be noted that in our definition of $\mathcal{M}$, $N_t$ refers to the total number of \emph{atoms} only, and we are not including the number of detected photons in our definition of the standard quantum limit. This could be considered `cheating', but we consider this a valid definition, as it is only the atoms that acquire the phase shift. Furthermore, in a typical atom interferometry experiment, atoms are by far the more valuable quantity: It is straight forward to increase the number of photons in our local oscillator, but difficult to increase the number of atoms in the atom interferometer. This scheme maximises the sensitivity of the atom interferometer \emph{for a given number of atoms}. This scheme can operate in a regime where $\Delta \phi$ is less than the true SQL of $1/\sqrt{N_t + N_p}$, where $N_p$ is the total number of detected photons, simply by reducing the gain factor $g$ to close to unity, such that there are less photons in the local oscillator. In this case, we achieved $\Delta \phi \approx 0.22/\sqrt{N_t} \approx 0.32/\sqrt{N_t+N_p}$  However, the absolute sensitivity is improved by increasing $g$, and as it will be simple to do so from an experimental stand point, we chose to perform our calculations in this regime. As $g$ becomes large, the fluctuations in $S_b$ become dominated by $\bhat_2$, and there is little benefit achieved by increasing $g$ further, as the limiting factor in reducing $V(S)$ becomes imperfect correlations between $\ahat_2$ and $\bhat_2$. 

\emph{Discussion} --- 
We can investigate if our  approximation of ignoring spontaneous scattering is valid by comparing the rate of scattering into $\ahat_2$, $\Gamma_{stim}$ to the total spontaneous  emission rate, $\Gamma_{spon}$. Assuming that the momentum width of the condensate is broad compared to the natural line width of the transition, then $\Gamma_{stim} \approx \mathcal{F}N_{seed}\Gamma_{spon}$, where $\mathcal{F}$ is the fraction of the total spontaneous emission distribution that results in a transition from $u_0(\boldr)e^{i\mathbf{k}_0\cdot \boldr}$ to $u_0(\boldr)e^{i(\mathbf{k}_0-\mathbf{k}_2)\cdot\boldr}$. Assuming a cigar shaped condensate with the long axis simultaneously aligned to $\mathbf{k}_2$, and the peak of the spontaneous emission dipole distribution, then $\mathcal{F} \approx \frac{3}{4\pi (k_2 \sigma)^2}$, where $\sigma$ is the characteristic width of the condensate.  For $^{87}$Rb in a trap with $1$ kHz radial confinement, $\mathcal{F} \approx 0.03$, and an initial seed of $10^4$ atoms in $|2\rangle$ is enough to ensure that the $\frac{\Gamma_{stim}}{\Gamma_{spon}} \approx 300$, and we can neglect the scattering into other modes. A feature of this scheme is that it is improved by having a larger BEC. The reason for this is that $\frac{\Gamma_{stim}}{\Gamma_{spon}}$ depends only on $N_{seed}$, but the maximum phase sensitivity is roughly $\mathcal{M} \sim\sqrt{2\sqrt{2}} \left(\frac{N_{seed}}{N_{t}}\right)^{1/4}$. Obviously a smaller seed is desirable, as the seed degrades the level of atom-light entanglement. It may be possible to enhance our scheme further by using strong coupling to a mode of an optical cavity rather than a seed to simulate scattering into a particular mode. Alternatively, seeding $\bhat_2$ with squeezed light to reduce fluctuations may also have a similar effect. 

The benefit of this scheme over other enhanced atom-interferometry schemes based on entanglement is that for a small increase in complexity of an atom-interferometry experiment, namely, employing homodyne detection on the scattered photons, a large gain in sensitivity can be achieved. Homodyne detection of Bragg-scattered photons with sufficient resolution to estimate the number of excitations generated in a BEC has recently been demonstrated \cite{pino_bragg}. Most atom-interferometry schemes benefit from a larger enclosed area, which requires high-momentum transfer beam splitters. Our scheme can be modified such that the difference in momentum between $\ahat_1$ and $\ahat_2$ is greater than $2\hbar k$ by using some other process to transfer momentum to one of the modes (such as state dependent Bloch oscillations  \cite{biraben_alpha}) after the Raman super-radiance beam-splitter. As long as this process doesn't exchange population between the two modes, the correlations aren't effected.

\emph{Acknowledgments} --- We would like to acknowledge useful discussions with Andy Ferris, Joe Hope, Nick Robins, Warwick Bowen, John Close, Matthew Davis, Joel Corney, and Holger Muller. This work was supported by the Australian Research Council Centre of Excellence for Quantum-Atom Optics, and by Australian Research Council discovery project DP0986893 and DE130100575.


\begin{thebibliography}{99}
\bibitem{chu99} A. Peters, K. Y. Chung, and S. Chu, Nature {\bf 400}, 849 (1999).
\bibitem{chu2001} A. Peters, K. Y. Chung, and S. Chu, Meteroliga, {\bf 38}, 25, (2001).
\bibitem{kasevich02} J. M. McGuirk, G. T. Foster, J. B. Fixler, M. J. Snadden, and M. A. Kasevich, Phys. Rev. A {\bf 65}, 033608 (2002).
\bibitem{kasevich97} T. L. Gustavson, P. Bouyer, and M. A. Kasevich, Phys. Rev. Lett. {\bf 78}, 2046 (1997).
\bibitem{stamperkurn07} M. Vengalattore, J. M. Higbie, S. R. Leslie, J. Guzman, L. E. Sadler, and D. M. Stamper-Kurn, Phys. Rev. Lett. {\bf 98}, 200801 (2007).
\bibitem{kasevichG} J. B. Fixler, G. T. Foster, J. M. McGuirk, and M. A. Kasevich, Science {\bf 315}, 5808 (2007).
\bibitem{biraben_alpha} R. Bouchendira, P. Clade, S. Guellati-Khelifa, F. Nez, F. Biraben, Phys. Rev. Lett. {\bf 106}, 080801 (2011).
\bibitem{close2011} J. E. Debs, P. A. Altin, T. H. Barter, D. D\"{o}ring, G. R. Dennis, G. McDonald, R. P. Anderson, J. D. Close, and N. P. Robins,  Phys. Rev. A {\bf 84}, 033610 (2011).
\bibitem{dowling} J. P. Dowling, Phys. Rev. A {\bf 57}, 4736 (1998).
\bibitem{ueda} M. Kitagawa and M. Ueda, Phys. Rev. A {\bf 47}, 5138 (1993).
\bibitem{hainekerr} S. A. Haine and M. T. Johnsson, Phys. Rev. A {\bf 80}, 023611 (2009). 
\bibitem{oberthaler2010} C. Gross, T. Zibold, E. Nicklas, J. Esteve, and M. K. Oberthaler, Nature {\bf 464}, 1165 (2010).
\bibitem{treutlein2010} M. F. Riedel, P. Bohl, Y. Li, T. W. Hansch, A. Sinatra, and P. Treutlein Nature {\bf 464}, 1170 (2010).
\bibitem{ertmer2011} B. L\"{u}cke, M. Scherer, J. Kruse, L. Pezze, F. Deuretzbacher, P. Hyllus, O. Topic, J. Peise, W. Ertmer, J. Arlt, L. Santos, A. Smerzi, C. Kelmpt, Science, {\bf 334}, 773, (2011).
\bibitem{squeezy1} S. A. Haine and J. J. Hope, Phys. Rev. A {\bf 72}, 033601 (2005).
\bibitem{squeezy3} S. A. Haine, M. K. Olsen, and J. J. Hope, Phys. Rev. Lett. {\bf 96}, 133601 (2006).
\bibitem{Schneble04} D. Schneble, G. K. Campbell, E. W. Streed, M. Boyd, D. E. Pritchard, W. Ketterle, Phys. Rev. A {\bf 69}, 041601 (2004)
\bibitem{Yoshikawa04} Y. Yoshikawa, T. Sugiura, Y. Torii, and T. Kuga, Phys. Rev. A {\bf 69}, 041603 (2004).
\bibitem{Cola04} M. M. Cola and N. Piovella, Phys. Rev. A {\bf 70}, 045601 (2004).
\bibitem{Wang05} T. Wang and S. F. Yelin, Phys. Rev. A {\bf 72}, 043804 (2005).
\bibitem{Meystre07} H. Uys and P. Meystre, Phys. Rev. A {\bf 75}, 033805 (2007).
\bibitem{Meystre2000} M. G. Moore, and P. Meystre, Phys. Rev. Lett. {\bf 85}, 5026 (2000).
\bibitem{Hilliard08} A. Hilliard, F. Kaminski, R. le Targat, C. Olausson, E. S. Polzik, and J. H. M\"{u}ller, Phys. Rev. A {\bf 78}, 051403 (2008).
\bibitem{Walls} D. F. Walls and G. J. Milburn, Quantum Optics (Berlin: Springer) (1994).
\bibitem{Steel98} M. J. Steel, M. K. Olsen, L. I. Plimak, P. D. Drummond, S. M. Tan, M. J. Collett, D.F. Walls, R. Graham, Phys. Rev. A {\bf 58}, 4824 (1998). 
\bibitem{Olsen09} M. K. Olsen and A. S. Bradley, Optics Communications {\bf 282}, 3924 (2009).
\bibitem{hainefwm} S. A. Haine and A. J. Ferris, Phys. Rev. A {\bf 84}, 043624 (2011).
\bibitem{pino_bragg} J. M. Pino, R. J. Wild, P. Makotyn, D. S. Jin, and E. A. Cornell, Phys. Rev. A {\bf 83}, 033615 (2011). 
\end{thebibliography}
\end{document}